\documentclass[twocolumn,showpacs,aps,prl,superscriptaddress,letterpaper]{revtex4}

\usepackage{dcolumn}
\usepackage{epsfig}

\RequirePackage{xspace}

\input babarsym

\providecommand{\btodspi}{\mbox{$B^0\to D_{s}^+\pi^-$}}
\providecommand{\btodsosk}{\mbox{$B^0\to D_{s}^{(*)-} K^+$}}
\providecommand{\btodsspi}{\mbox{$B^0\to D_{s}^{*+}\pi^-$}}
\providecommand{\btodsk}{\mbox{$B^0\to D_{s}^{-}K^+$}}
\providecommand{\btodssk}{\mbox{$B^0\to D_{s}^{*-}K^+$}}
\providecommand{\btodsospi}{\mbox{$B^0\to D_{s}^{(*)+}\pi^-$}}

\providecommand{\btodospi}{\ensuremath{B^0\to D^{(*)-}\pi^+}}

\providecommand{\dsphipi}{\mbox{$D^+_{s}\to \phi\pi^+$}}

\providecommand{\De}{\ensuremath{\Delta E}}
\providecommand{\mes}{\ensuremath{m_{ES}}}

\providecommand{\mDs}{\ensuremath{m(D_s)}}

\providecommand{\stwobg}{\ensuremath{\sin(2\beta+\gamma)}\xspace}

\def\journal#1#2#3#4{#1~{\bf #2}, #3 (#4)}
\def\PL#1#2#3{\journal{Phys.\ Lett.}{#1}{#2}{#3}}

\def\PR#1#2#3{\journal{Phys.\ Rev.}{#1}{#2}{#3}}
\def\PRL#1#2#3{\journal{Phys.\ Rev. Lett.}{#1}{#2}{#3}}
\def\ZP#1#2#3{\journal{Z.\ Phys.}{#1}{#2}{#3}}

\newcommand{\etal}{{\em et al.}}

\newcommand{\lumi}{\ensuremath{230\times 10^6}}
\newcommand{\brdspiOnly}{\ensuremath{1.3\pm 0.3\ \stat\ \pm 0.2\ \syst}}
\newcommand{\brdsspiOnly}{\ensuremath{2.8\pm 0.6\ \stat\ \pm 0.5\ \syst}}
\newcommand{\brdskOnly}{\ensuremath{2.5\pm 0.4\ \stat\ \pm 0.4\ \syst}}
\newcommand{\brdsskOnly}{\ensuremath{2.0\pm 0.5\ \stat\ \pm 0.4\ \syst}}
\newcommand{\brdspi}{\ensuremath{\BR(\btodspi)=(\brdspiOnly})\times 10^{-5}}
\newcommand{\brdsspi}{\ensuremath{\BR(\btodsspi)=(\brdsspiOnly)\times 10^{-5}}}
\newcommand{\brdsk}{\ensuremath{\BR(\btodsk)=(\brdskOnly)\times 10^{-5}}}
\newcommand{\brdssk}{\ensuremath{\BR(\btodssk)=(\brdsskOnly)\times 10^{-5}}}

\newcommand{\BABARPubYear}    {06}

\newcommand{\BABARPubNumber}  {026}

\newcommand{\SLACPubNumber} {11803}

\begin{document}

\begin{flushright}
\small
\babar-PUB-\BABARPubYear/\BABARPubNumber\\
SLAC-PUB-\SLACPubNumber\\
\end{flushright}
\vspace{-0.5in}

\title{
{\large \bf Observation of Decays \btodsospi\ and \btodsosk } 
}

%
\author{B.~Aubert}
\author{R.~Barate}
\author{M.~Bona}
\author{D.~Boutigny}
\author{F.~Couderc}
\author{Y.~Karyotakis}
\author{J.~P.~Lees}
\author{V.~Poireau}
\author{V.~Tisserand}
\author{A.~Zghiche}
\affiliation{Laboratoire de Physique des Particules, F-74941 Annecy-le-Vieux, France }
% ...
\collaboration{The \babar\ Collaboration}
\noaffiliation

\date{April 5, 2006}

\begin{abstract}
We report the observation of decays
\btodsospi\ and \btodsosk\ in a 
sample of \lumi\ $\FourS\to\BB$ events recorded
with the \babar\ detector 
at the PEP-II asymmetric-energy $e^{+}e^{-}$ storage ring.
We measure the branching fractions $\brdspi$, $\brdsk$, $\brdsspi$, and
$\brdssk$. The significance of the measurements to differ from zero
are 5, 9, 6, and 5 standard deviations, respectively. 
\end{abstract}


\maketitle

Within the Cabibbo-Kobayashi-Maskawa (CKM) model of quark-flavor
mixing~\cite{CKM},  
\CP\ violation manifests itself as a non-zero area of the unitarity
triangle~\cite{Jarlskog}. 
One of the important experimental tests of the model is
the determination of the angle 
$\gamma = {\rm arg}(-V_{ud}V_{ub}^{*}/V_{cd}V_{cb}^{*})$ 
of the unitarity triangle. A
measurement of 
\stwobg\ can be obtained from the study 
of the time dependence of the $\Bz,\Bzb{\to} D^{(*)-} \pi^+$~\cite{chconj}
decay rates, and specifically of the interference between the
CKM-favored \Bz\ decay amplitude and CKM-suppressed \Bzb\
amplitude~\cite{sin2bg}.  
The first
measurements of the \CP\ asymmetry in decays $\Bz{\to}
D^{(*)\mp}\pi^\pm$ have recently been published~\cite{ref:s2bgDPi}. 

The measurement of \stwobg\ 
in
$\Bz{\to} D^{(*)\mp}\pi^\pm$ decays 
requires knowledge of the ratios of the decay amplitudes, 
$r({D^{(*)}\pi})=|A(\Bz{\to} D^{(*)+}\pi^-)/A(\Bz{\to}D^{(*)-}\pi^+)|$. 
However,
direct measurement of the branching fractions
$\BR(\Bz{\to} D^{(*)+}\pi^-)$ is not 
possible with the currently available data sample due to the presence
of the overwhelming background from $\Bzb{\to} D^{(*)+}\pi^-$.
However, assuming SU(3) flavor symmetry, $r({D^{(*)}\pi})$ can be related
to the branching fraction (BF) of the decay \btodsospi~\cite{sin2bg}:
\begin{equation}
r({D^{(*)}\pi}) = 
  \tan\theta_c\,
  \frac{f_{D^{(*)}}}{f_{D^{(*)}_s}}\sqrt{\frac{\BR(\btodsospi)}{\BR(\btodospi)}}
  \ ,
\label{eq:rDPi}
\end{equation}
where $\theta_c$ is the Cabibbo angle, and $f_{D^{(*)}}/f_{D^{(*)}_s}$
is the ratio of $D^{(*)}$ and $D^{(*)}_s$ meson decay
constants~\cite{fdsd}. Other 
SU(3)-breaking effects are believed to affect $r({D^{(*)}\pi})$ by
less than $30\%$~\cite{ref:s2bgDPi}.

\begin{figure}[h]
\begin{center}
\epsfig{file=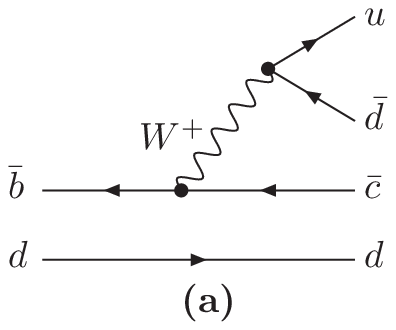,width=1in} 
\epsfig{file=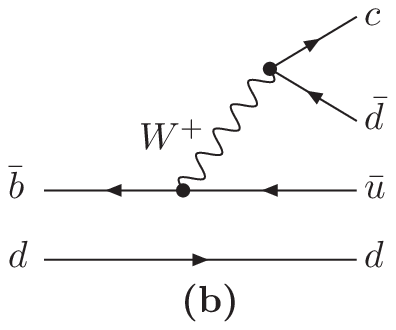,width=1in} 
\epsfig{file=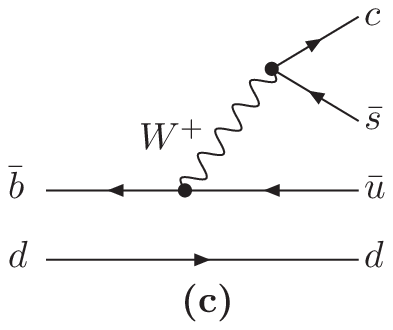,width=1in}\\
\epsfig{file=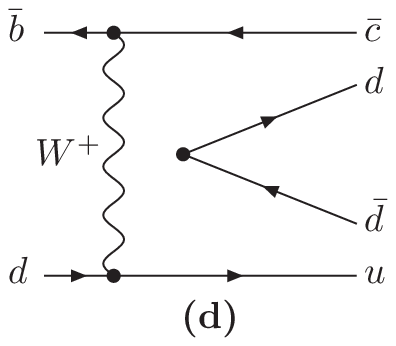,width=1in} 
\epsfig{file=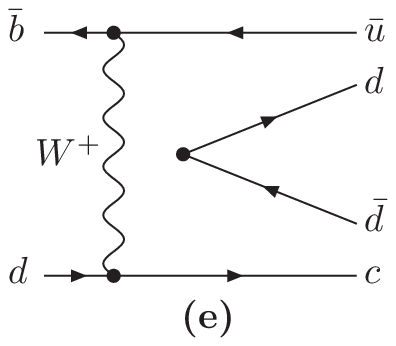,width=1in}
\epsfig{file=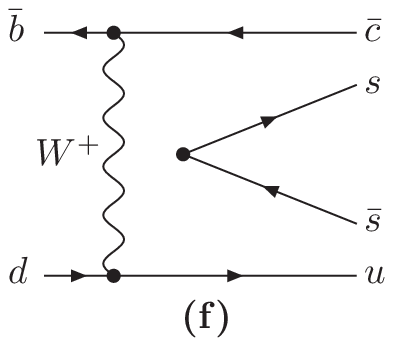,width=1in}
\end{center}
\vspace{-0.5cm}
\caption{Dominant Feynman diagrams for (a) CKM-favored decay
$\Bz\to D^{(*)-}\pi^+$, (b) doubly CKM-suppressed decay
$\Bz\to D^{(*)+}\pi^-$, and (c) the SU(3) flavor symmetry related decays 
\btodsospi;
(d) the color-suppressed $W$-exchange contributions
  to $\Bz\to D^{(*)-}\pi^+$, (e) $\Bz\to D^{(*)+}\pi^-$, and
(f) decay \btodsosk.}    
\label{fig:diag} 
\end{figure}

Since  \btodsospi\ has four
different quark flavors in the final state, only a single amplitude
contributes to the decay (Fig.~\ref{fig:diag}c). 
On the other hand, there are two diagrams contributing to 
$\Bz\to D^{(*)-}\pi^+$ and $\Bz\to D^{(*)+}\pi^-$:  
tree amplitudes (Fig.~\ref{fig:diag}a,b) and color-suppressed direct
$W$-exchange amplitudes (Fig.~\ref{fig:diag}d,e). The latter are
assumed to be negligibly small in Eq.~(\ref{eq:rDPi}). The decays \btodsosk\
(Fig.~\ref{fig:diag}f) probe the size of the $W$-exchange
amplitudes relative to the dominant processes $\Bz\to
D^{(*)-}\pi^+$. The rate of \btodsosk\
decays could be enhanced by final state rescattering~\cite{Wexch}, in
addition to the $W$-exchange amplitude. 

The branching fractions $\BR(\btodspi)$ and $\BR(\btodsk)$ have been
measured previously 
by the \babar~\cite{priorBaBar} and Belle~\cite{priorBelle}
collaborations, but the decays \btodsspi\ and \btodssk\ have never
been observed. 
In this Letter we present  new measurements of the
decays  \btodsospi\  and \btodsosk.
The analysis uses  a sample  of \lumi\ \FourS\ decays into \BB\ 
 pairs 
 collected  with the \babar\ detector
at the \pep2\ asymmetric-energy B factory~\cite{pep}.

Since the \babar\ detector is described in detail
elsewhere~\cite{detector},  
only the components that are crucial to this analysis are
 summarized here. 
Charged particle tracking is provided by a five-layer double-sided silicon
vertex tracker (SVT) and a 40-layer drift chamber (DCH). 
Ionization energy loss ($dE/dx$) in
the DCH and SVT and Cherenkov radiation detected in a ring-imaging
device are used for charged-particle identification. 
Photons are identified and measured using
the electromagnetic calorimeter (EMC), which is comprised of 6580
thallium-doped CsI 
crystals. These systems are mounted inside a 1.5 T solenoidal
superconducting magnet. 
We use the GEANT4~\cite{geant} software to simulate interactions of particles
traversing the \babar\ detector, taking into account the varying
detector conditions and beam backgrounds. 

We pre-select events which have a minimum of four reconstructed
charged tracks and 
a total measured energy greater than $4.5$~GeV, determined
using all charged tracks and neutral clusters with energy above 30 MeV.
In order to reduce ``continuum'' $\epem\to q\bar{q}\ (q=u,d,s,c)$
background, the ratio of the second 
and zeroth order Fox-Wolfram moments~\cite{fox} 
must be less than 0.5.

Candidates for \Ds\ mesons 
are reconstructed in the modes
$\Ds\to\phi \pi^+$, $\KS K^+$ and $\Kstarzb\Kp$, 
with $\phi{\to} K^+K^-$, $\KS {\to} \pip \pim$, and $\Kstarzb{\to} K^-\pi^+$. 
%
%
The $\KS$ candidates are reconstructed from two
oppositely-charged tracks, and their momenta are required to make an angle 
$|\theta_\mathrm{flight}|<11^{\circ}$ with the line connecting their
vertex and $\epem$ interaction point. All other tracks are  
required  to originate from the $\epem$ interaction region.
In order to reject background from $\Dp{\to}\KS\pip$ or $\Kstarzb\pip$,
the $\Kp$ candidate in the reconstruction of $\Ds{\to}\KS\Kp$ or
$\Kstarzb\Kp$ is 
required to satisfy positive kaon identification criteria with an
efficiency of 85\%  and 5\% pion misidentification probability. The same 
selection is used to identify kaon daughters of the
$B$ mesons in decays \btodsosk. In all other cases, kaons are
not positively identified, but instead candidates passing pion
selection are rejected. Such ``pion veto'' has an efficiency of 95\%
for kaons and  20\%  for pions. 
Pion daughters of $B$ mesons in the decays \btodsospi\ are required
to be positively identified. 
Decay products of $\phi$, $\Kstarzb$, $\KS$, $\Ds$, and $\Bz$ candidates are 
constrained to originate from a single vertex. 

We reconstruct  \Dss\ candidates in the mode  $\Dss{\to}\Ds\gamma$ by
combining  \Ds\ and photon candidates.
Photon candidates are required to be consistent with an electromagnetic
shower in the EMC, and have an energy greater than $100$~\mev in the
laboratory frame. 
When forming a \Dss, the \Ds\ candidate is required to have invariant mass
within 10~\mevcc\ of the nominal value~\cite{PDG2004}. 

%
%
After an initial pre-selection, we identify candidates for \btodsospi\ and
\btodsosk\ using a likelihood ratio 
$R_L =
\mathcal{L}_\mathrm{sig}/(\mathcal{L}_\mathrm{sig}+\mathcal{L}_\mathrm{bkg})$,
where
$\mathcal{L}_\mathrm{sig}=\prod_{i}\mathcal{P}_{\mathrm{sig}}(x_i)$
is the multivariate likelihood for 
signal events and
$\mathcal{L}_\mathrm{bkg}=\prod_{i}\mathcal{P}_{\mathrm{bkg}}(x_i)$
is the likelihood for 
background events. The ratio $R_L$ has a maximum at $R_L=1$ for
signal events, and at $R_L=0$ for background originating
from continuum events. It also discriminates
well against generic $B$ decays without a real $\Ds$ meson in the final
state. 
The likelihoods for signal and background events are
computed as a product of the probability density functions (PDFs)
$\mathcal{P}_{\mathrm{sig}}(x_i)$ and $\mathcal{P}_{\mathrm{bkg}}(x_i)$ for a
number of selection variables $x_i$: invariant masses of the $\phi$,
$\Kstarzb$ and $\KS$ candidates, $\chi^2$ confidence level of the vertex
fit for the $\Bz$ and $\Ds$ mesons, the helicity angles of the $\phi$,
$\Kstarzb$, and $\Dss$ meson decays, the mass difference
$\Delta m(\Dss) = m(\Dss)-m(\Ds)$, the polar angle $\theta_B$ of the
$B$ candidate 
momentum vector with respect to the beam axis in the
$\epem$ center-of-mass (c.m.) frame, the angle $\theta_{T}$ 
between the thrust axis of  the  $B$ candidate and the thrust
axis of all other particles in the event in c.m. frame, 
and event topology variable $\mathcal{F}$. 
Correlations among these variables are small. 
The helicity angle
$\theta_H$ is defined as the angle between one of the decay products  of
a vector meson and the flight direction of its parent particle, in the meson's
rest frame. Polarization of the vector mesons
in the signal decays causes their helicity angles to be distributed as
$\cos^2\theta_H$ ($\phi$ and $\Kstarzb$) or $\sin^2\theta_H$ (\Dss),
while the random background combinations tend to 
produce a more uniform distribution in $\cos\theta_H$. 

Variables $\cos\theta_B$, $\cos\theta_T$, and $\mathcal{F}$
discriminate between spherically-symmetric \BB\ events and jetty
continuum background using event topology. 
\BB\ pairs form a nearly uniform $|\cos\theta_T|$
distribution, while $|\cos\theta_T|$ distribution for the continuum
peaks at 1.
A linear (Fisher)
discriminant $\mathcal{F}$ is
derived from the values of sphericity and thrust for the event, 
and the two Legendre moments $L_0$ and
$L_2$ of the energy flow 
around the $B$-candidate thrust axis~\cite{ref:legendre}.
Finally, the polar angle $\theta_B$ is distributed as $\sin^2\theta_B$
for real $B$ decays, while being nearly flat in $\cos\theta_B$ for
the continuum.

We select \btodspi\ and \btodsk\ candidates that satisfy
$R_L>0.75$, and accept \btodsspi\ and \btodssk\ candidates with
$R_L>0.8$. We measure the relative efficiency $\varepsilon_{R_L}$ of
the $R_L$ selection in a copious 
data sample of decays $\Bz\to D^-\pi^+$ ($D^-\to
K^+\pi^-\pi^-,\, \KS\pi^-$) 
and $B^+\to\Dbar^{*0}\pi^+$ ($\Dbar^{*0}\to
\Dbar^0\gamma,\, D^0\to K^-\pi^+$) 
in which the kinematics is similar to that of our signal events, and find
that it is consistent with Monte Carlo estimates 
$\varepsilon_{R_L}\approx 70\%$. The fraction of continuum
background events passing the selection varies between $2\%$ and $15\%$,
depending on the mode.

%
%
We identify the signal using the invariant mass \mDs\ of $D_s$
candidates and two kinematic variables \mes\ and
$\De$.  The first is the beam-energy-substituted mass 
$\mes = \sqrt{ (s/2 +
  \mathbf{p}_{i}\cdot\mathbf{p}_{B})^{2}/E_{i}^{2}- 
  \mathbf{p}^{2}_{B}}$, 
where $\sqrt{s}$ is the total c.m.
energy,  $(E_{i},\mathbf{p}_{i})$ is the four-momentum of the initial
\epem\ system and $\mathbf{p}_{B}$ is the \Bz\ candidate momentum,
both measured in the laboratory frame. The second variable is 
$\De = E^{*}_{B} - \sqrt{s}/2$, where $E^{*}_{B}$ is the \Bz\ candidate
energy in the c.m. frame. 
For signal events, the \mes\ distribution is gaussian centered at the $B$ meson
mass with a resolution of about $2.5$~\mevcc, and the \De\ distribution has a
maximum near zero with a resolution of about 17~\mev. The invariant
mass \mDs\ has a resolution of
$(5-6)$~\mevcc, depending on the $\Ds$ decay mode. We define a fit region
$5.2<\mes<5.3$~\gevcc, $|\De|<36$~\mev, and 
$|\mDs-\mDs_\mathrm{PDG}|<50$~\mevcc\ for \btodspi\ and
\btodsk\ candidates, where 
$\mDs_\mathrm{PDG}$ is the world average $D_s$ mass~\cite{PDG2004}. 
For \btodsspi\ and \btodssk, we require
$|\mDs-\mDs_\mathrm{PDG}|<10$~\mevcc. 

Less than 20\% of the selected events in the \btodsspi\ and \btodssk\
channels ($<4\%$ in \btodspi\ and \btodsk)
contain two or more 
candidates that satisfy the criteria listed above. In
such events we select a single $B^0$ candidate based on
an event $\chi^2$ formed with \mDs\ and $\Delta m(\Dss)$ and their
uncertainties, and the \De\
variable. Such selection does not bias background distributions
significantly.

Four classes of background contribute to the fit region. 
First is the {\em combinatorial background\/}, in which a true or fake
$D_s^{(*)}$ candidate is combined with a randomly-selected pion or
kaon. 
Second,  $B$ meson decays such as  $\Bzb{\to}D^{(*)+}\pim, \rho^-$ with
$\Dp{\to}\KS\pip$ or $\Kstarzb\pip$ can 
constitute a background for  the \btodsospi\ modes if
 the pion in the $D$ decay is misidentified as a kaon ({\it{reflection
background}}). The reflection background has nearly the same \mes\ distribution
as the signal but different distributions in \De\ and \mDs.
The corresponding backgrounds for the
\btodsk\ mode ($\Bz{\to}\Dm K^{(*)+}$) are negligible. 
Third, rare $B$  decays into the same final state, such as
$\Bz{\to} \Kbar^{(*)0}\Kp\pim$ or $\Kbar^{(*)0}\Kp\Km$
({\it{charmless background}}),
have the same \mes\ and \De\ distributions as the \btodspi\ or
\btodsk\ signal, but are nearly flat in \mDs. The charmless background is
significant in \btodspi\ and \btodsk\ decays, but is negligible for
\btodsspi\ and \btodssk. Finally, 
{\em crossfeed background\/} from misidentification of 
$\Bbar^0\to D_s^{(*)-}\pi^+$ events
as \btodsosk\ signal, and vice versa, needs to be taken into account. 

%
%
We perform a two-dimensional unbinned extended 
maximum-likelihood fit to the \mes\ and \mDs\ distributions
to extract $\BR(\btodspi)$ and $\BR(\btodsk)$ and constrain the
contributions from charmless background modes. 
Charmless backgrounds are negligible for \btodsspi\ and \btodssk, 
and we determine the BFs of these decays with a
one-dimensional fit to the \mes\ distribution. 
For each $B$ decay, we simultaneously fit distributions in three 
\Ds\ decay modes,  
constraining the signal BFs to a common value. 
The likelihood function contains the contributions 
of the signal and the four background components discussed above. 
The combinatorial background is 
described in \mes\ by a threshold function~\cite{argus}, 
$dN/dx\propto x\sqrt{1-2x^{2}/s}\exp\left[-\xi\left(1-2x^{2}/s\right)\right]$.
In \mDs, the combinatorial background is well described by a
combination of a first-order polynomial (fake \Ds\ candidates) and a
gaussian with $(5-6)$~\mevcc\ resolution (true \Ds\
candidates). The charmless background is parameterized by the signal
gaussian shape in \mes\ and a first order polynomial in \mDs. 

\begin{figure}
\begin{center}
\epsfig{file=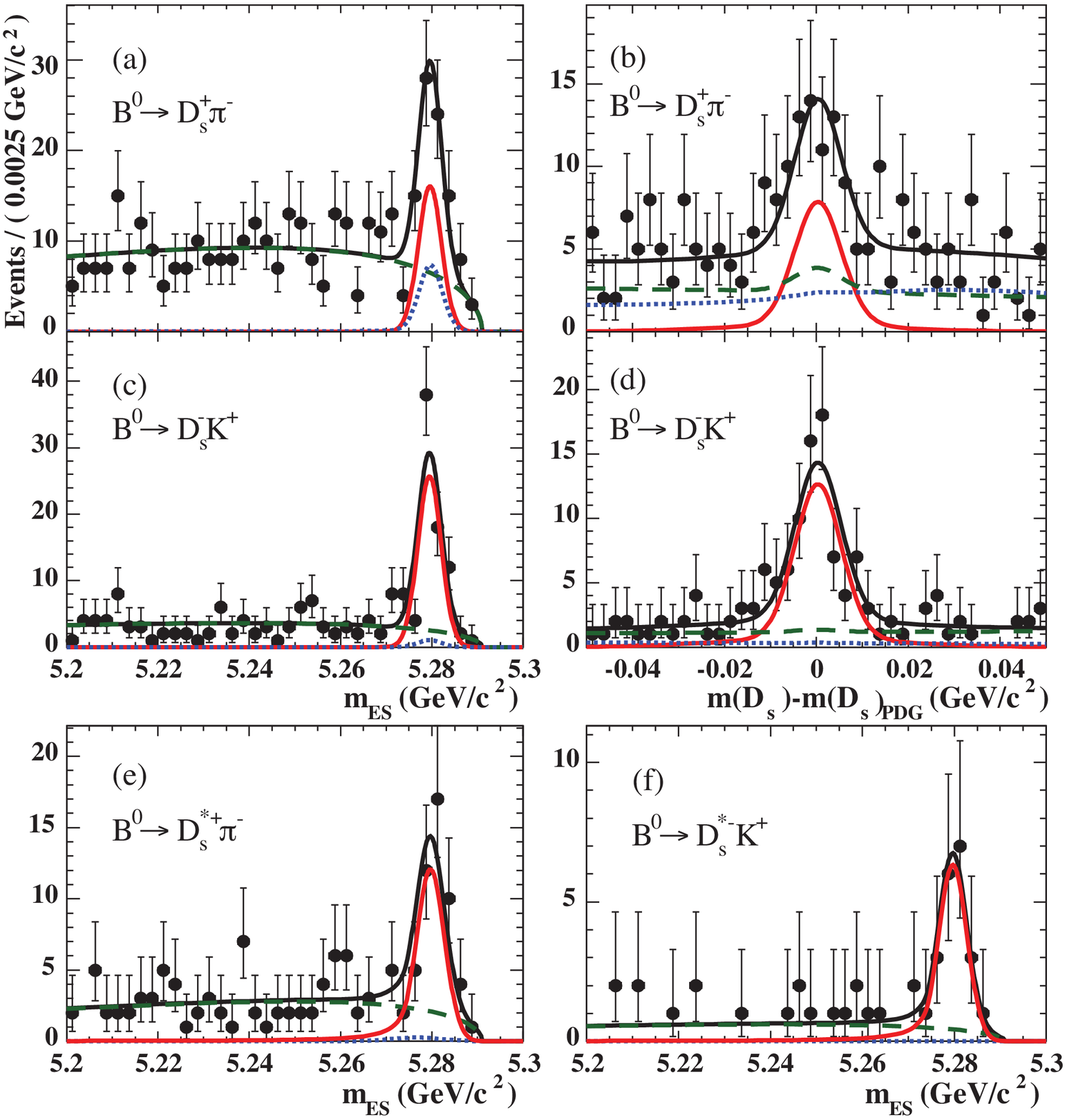,width=\columnwidth}
\end{center}
\vspace{-0.8cm}
\caption{ (a,c,e,f) \mes\ projection of the fit with 
$|m(\Ds)-m(\Ds)_\mathrm{PDG}|<10$~\mevcc and 
(b,d) \mDs\ projection with $5.275<\mes<5.285$~\gev for (a,b) \btodspi,
(c,d) \btodsk, (e) \btodsspi, and (f) \btodssk. The black solid curve
corresponds to the full PDF from the combined
fit to all \Ds\ decay modes. Individual contributions are shown 
as solid red (signal PDF), green dashed (combinatorial background),
and blue dotted (sum of reflection, charmless, and crossfeed
backgrounds) curves. 
}
\label{fig:fit} 
\end{figure}

For \btodspi\ and \btodsk\ decays, the fit constrains 
14 free parameters: the shape of the combinatorial background
$\xi$ (1 parameter for all \Ds\ modes), the slope of the combinatorial
and charmless backgrounds in \mDs\ (3 parameters), the fraction of
true \Ds\ candidates in 
combinatorial background (3), the number of combinatorial background
events (3), the number of charmless events (3), and the BF
of the signal mode (1). 
The signal yields for each \Ds\ mode are
expressed as 
$N_{\mathrm{sig}\, i} = N_{\BB}\, \BR_{\mathrm{sig}}\, \BR_i\, \varepsilon_i$, 
where $N_{\BB}=\lumi$,
$\BR_i$ is the \Ds\ BF for the mode, $\varepsilon_i$ is
the reconstruction efficiency, and $\BR_{\mathrm{sig}}$ is the 
BF (fit parameter) for the decay. 
For the \btodsspi\ and \btodssk\ decays, 5 free
parameters are determined by the fit: $\xi$ (1 parameter for all \Ds\
modes), the number of combinatorial background events (3), and the
BF of the signal mode (1). 
The BFs of the channels contributing to the reflection
background  are fixed in the
fit to the current world average values~\cite{PDG2004}, and the
BFs of the crossfeed backgrounds are determined by
iterating the fits over each $B$ decay mode. 
The results of the fits are shown in Fig.~\ref{fig:fit} and summarized
in Table~\ref{tab:fit}.

%

\begin{table*}[t!]
\caption{The number of reconstructed candidates ($N_\mathrm{raw}$), 
the  signal yield ($N_\mathrm{sig}$), computed from the fitted
branching fractions, combinatorial background
($N_\mathrm{comb}$), and the sum of charmless, reflection, and
crossfeed contributions ($N_\mathrm{peak}$), 
extracted from the likelihood fit. 
Also given are 
the reconstruction efficiency ($\varepsilon$), 
 the probability ($P_\mathrm{bkg}$) of the data being consistent with 
the background in the absence of signal, and 
the measured branching fraction \BR. 
The first uncertainty is  statistical, and the second is systematic. 
}

\begin{center}
\begin{tabular*}{\textwidth}{@{\extracolsep{\fill}} ll c c c c c c c c c c c c} \hline \hline
$B$ mode& $D_s$ mode & $N_\mathrm{raw}$ & $N_\mathrm{sig}$  & $N_\mathrm{comb}$ & 
$N_\mathrm{peak}$ &
$\varepsilon$(\%)&  $P_\mathrm{bkg}$ & \BR ($10^{-5}$) & $\BR\times\BR(\dsphipi)$ \\
&& & & & & & & & $(10^{-6})$ \\
\hline
           & $\Ds{\to}\phi\pip$    & $405$ &$21\pm5$ &$364\pm20$ &$21\pm8$  &$29.3$ & & & \\
$\btodspi$ & $\Ds{\to}\Kstarzb\Kp$ & $677$ &$16\pm4$ &$604\pm26$ &$58\pm12$ &$20.0$ &$3\cdot10^{-6}$ & $1.3\pm0.3\pm0.2$ & $0.63\pm0.15\pm0.05$ \\
           & $\Ds{\to}\KS\Kp$      & $223$ &$11\pm3$ &$197\pm15$ &$16\pm6$  &$22.1$ & & & \\
\hline
           & $\Ds{\to}\phi\pip$    & $46$ &$18\pm4$ &$29\pm6$ & $0$ &$13.0$ & & & \\
$\btodsspi$&$\Ds{\to}\Kstarzb\Kp$  & $67$ &$14\pm3$ &$48\pm8$ & $1$ &$8.9$ & $3\cdot10^{-8}$& $2.8\pm0.6\pm0.5$ & $1.32\pm0.27\pm0.15$ \\
           & $\Ds{\to}\KS\Kp$      & $19$ &$10\pm2$ &$12\pm4$ & $1$ &$9.6$ & & & \\
\hline
          & $\Ds{\to}\phi\pip$    & $197$ &$32\pm5$ &$151\pm13$ &$8\pm6$  &$23.4$ & 
& & \\
$\btodsk$ & $\Ds{\to}\Kstarzb\Kp$ & $331$ &$27\pm4$ &$306\pm18$ &$-4\pm6$ &$17.6$ & $3\cdot10^{-19}$& $2.5\pm0.4\pm0.4$ & $1.21\pm0.17\pm0.11$ \\
          & $\Ds{\to}\KS\Kp$      & $101$ &$18\pm3$ &$82\pm10$  &$9\pm5$  &$19.0$ & 
& & \\
\hline
           & $\Ds{\to}\phi\pip$    & $15$ &$9\pm2$ &$8\pm3$ & - & $8.9$  &&&\\
$\btodssk$ & $\Ds{\to}\Kstarzb\Kp$ & $16$ &$8\pm2$ &$7\pm3$ & - & $6.6$ &$2\cdot10^{-5}$& $2.0\pm0.5\pm0.4$ & $0.97\pm0.24\pm0.12$ \\
           & $\Ds{\to}\KS\Kp$      & $10$ &$5\pm1$ &$5\pm3$ & - & $6.7$  &&&\\
\hline\hline
\end{tabular*}
\end{center}
\label{tab:fit}
\end{table*}

%
The systematic errors are dominated by the 13\% relative uncertainty for
\BR (\Ds$\rightarrow\phi\pip$)~\cite{BRDsPhiPi}. The uncertainties in
the relative BFs
$\BR(\Ds{\to}\Kstarzb\Kp)/\BR(\Ds{\to}\phi\pip)$ and 
$\BR(\Ds{\to}\KS\Kp)/\BR(\Ds{\to}\phi\pip)$ contribute $(5-7)\%$,
depending on the decay channel. Uncertainties in the selection
efficiency are estimated to be $3\%$ for \btodspi and \btodsk, and
$7\%$ for \btodsspi and \btodssk. The uncertainties in the reflection
and crossfeed backgrounds are below 1\% for all decay channels. 
The rest of the systematic errors, which include the uncertainties in
tracking, photon and \KS\ reconstruction, charged-kaon identification 
efficiencies, and variations of the PDF shapes between data and Monte
Carlo, amount to $(6-7)\%$. 

The ratio $P_\mathrm{bkg} = \mathcal{L}_0/\mathcal{L}_{\max}$, where 
$\mathcal{L}_{\max}$ is the maximum likelihood value, and 
$\mathcal{L}_0$ is the likelihood for a fit with the signal
contribution set to zero, describes the probability of 
the background to fluctuate to the observed number of events. 
Including systematic uncertainties and assuming gaussian-distributed
errors, it corresponds to the significance of signal 
observation of 
$5$ (\btodspi), $6$ (\btodsspi), $9$ (\btodsk), and
$5$ (\btodssk) standard deviations. 
This is the first observation of \btodspi,
\btodsspi, and \btodssk\ decays. 

The BF results are collected in Table~\ref{tab:fit}.
Since the dominant uncertainty 
comes from the knowledge of the \Ds\ BFs, we also report
the products $\BR\times \BR(\Ds\rightarrow\phi\pip)$. 
The BFs for \btodsosk\ are small
compared to the dominant decays $\Bz\to D^{(*)-}\pi^+$, implying
relatively insignificant contributions from the color-suppressed
$W$-exchange diagrams. 
Assuming SU(3) relation, Eq.~(\ref{eq:rDPi}), we determine 
$r({D\pi}) = (1.3\pm0.2\stat\pm0.1\syst)\times10^{-2}$, and 
$r({D^{*}\pi}) = (1.9\pm0.2\stat\pm0.2\syst)\times10^{-2}$, 
which implies small $C\! P$ asymmetries in
$\Bz{\to} D^{(*)\mp}\pi^\pm$ decays. These results
supersede our previously published measurements~\cite{priorBaBar}.

We are grateful for the excellent luminosity and machine conditions
provided by our \pep2\ colleagues, 
and for the substantial dedicated effort from
the computing organizations that support \babar.
The collaborating institutions wish to thank 
SLAC for its support and kind hospitality. 
This work is supported by
DOE
and NSF (USA),
NSERC (Canada),
IHEP (China),
CEA and
CNRS-IN2P3
(France),
BMBF and DFG
(Germany),
INFN (Italy),
FOM (The Netherlands),
NFR (Norway),
MIST (Russia), and
PPARC (United Kingdom). 
Individuals have received support from CONACyT (Mexico), 
Marie Curie EIF (European Union),
the A.~P.~Sloan Foundation, 
the Research Corporation,
and the Alexander von Humboldt Foundation.


\begin{thebibliography}{99}
\bibitem{chconj}
Charge conjugation is implied throughout this letter, unless 
explicitly stated.
\end{thebibliography}
\end{document}